\documentclass[%
 reprint,
superscriptaddress,
 amsmath,amssymb,
 aps,
floatfix,
]{revtex4-2}

\usepackage[dvips]{graphicx}
\usepackage{dcolumn}
\usepackage{bm}
\usepackage{amsmath}
\usepackage{xcolor}
\usepackage{float}
\usepackage{multirow}
\usepackage[utf8]{inputenc}
\usepackage[T1]{fontenc}
\usepackage[english]{babel}
\draft 
	
\begin{document}

\title{
Disentangling the Sub-Cycle Electron Momentum Spectrum in Strong-Field Ionization
}

\author{Nicholas Werby}
\affiliation{Stanford PULSE Institute, SLAC National Accelerator Laboratory\\
2575 Sand Hill Road, Menlo Park, CA 94025}
\affiliation{Department of Physics, Stanford University, Stanford, CA 94305}

\author{Adi Natan}
\affiliation{Stanford PULSE Institute, SLAC National Accelerator Laboratory\\
2575 Sand Hill Road, Menlo Park, CA 94025}
\affiliation{Department of Physics, Stanford University, Stanford, CA 94305}

\author{Ruaridh Forbes}
\affiliation{Stanford PULSE Institute, SLAC National Accelerator Laboratory\\
2575 Sand Hill Road, Menlo Park, CA 94025}
\affiliation{Department of Physics, Stanford University, Stanford, CA 94305}

\author{Philip H. Bucksbaum}
\affiliation{Stanford PULSE Institute, SLAC National Accelerator Laboratory\\
2575 Sand Hill Road, Menlo Park, CA 94025}
\affiliation{Department of Physics, Stanford University, Stanford, CA 94305}
\affiliation{Department of Applied Physics, Stanford University, Stanford, CA 94305}

\date{\today}

\begin{abstract}
Quantum calculations of tunneling in strong-field ionization (SFI) predict intricate momentum distributions due to sub-laser-cycle attosecond electron dynamics. These are obscured in most experiments by the dominance of inter-cycle interference patterns which are the hallmark of above-threshold ionization (ATI).
Highly controlled 1- to 2-cycle laser pulses produce less inter-cycle interference but cannot accurately recreate the sub-cycle features produced by uniform cycle calculations due to the effect of the carrier envelope of the pulse. 
We present a simple and effective technique to recover these sub-cycle features in experimental multi-cycle spectra. We time-filter the momentum distribution to highlight features originating from the interference of electron trajectory pairs with ionization times less than one field cycle apart.
This method removes the ATI patterns and reveals sub-cycle interference structures in unprecedented detail. We can resolve new modulations in holographic structures that have not been previously noted in earlier experiments and which provide a new reference for comparing to calculations.  

\end{abstract}

\maketitle

When an atom or molecule is photoionized by a strong laser field, the photoelectron undergoes a series of complex field-driven dynamics \cite{okino_direct_2015,smirnova_strong-field_2009,xu_self-imaging_2010,willenberg_holographic_2019,tan_time-resolving_2019,he_direct_2018,walt_dynamics_2017,zhou_near-forward_2016,meckel_signatures_2014,marchenko_criteria_2011}. Electron vector momentum distributions obtained using angle-resolving techniques, such as velocity map imaging (VMI), contain detailed information on rapidly evolving molecular geometries of the parent \cite{vu_dynamic_2017, krecinic_multiple-orbital_2018, morishita_retrieval_2009}, holographic structures from the interference of electron trajectories \cite{huismans_time-resolved_2011,tan_time-resolving_2019,walt_dynamics_2017,willenberg_holographic_2019,he_direct_2018,marchenko_criteria_2011, faria_it_2020, maxwell_coulomb-free_2018,haertelt_probing_2016,porat_attosecond_2018}, and nonlinear electron interactions with the strong field \cite{blaga_strong-field_2009}. Each of these processes can produce patterns in the spectrum with characteristic features; however, patterns from different processes usually overlap on the electron detector, which impedes straightforward analysis. Experiments in the strong-field ionization (SFI) regime, where the ionization during each field cycle is significant, must either have carefully chosen running parameters such as ellipticity or field shape to emphasize specific dynamics, or significant analysis must be performed on the observed spectrum to disentangle these patterns to make the desired measurement. 

\begin{figure*}
	\centering
	\includegraphics[width=1\linewidth]{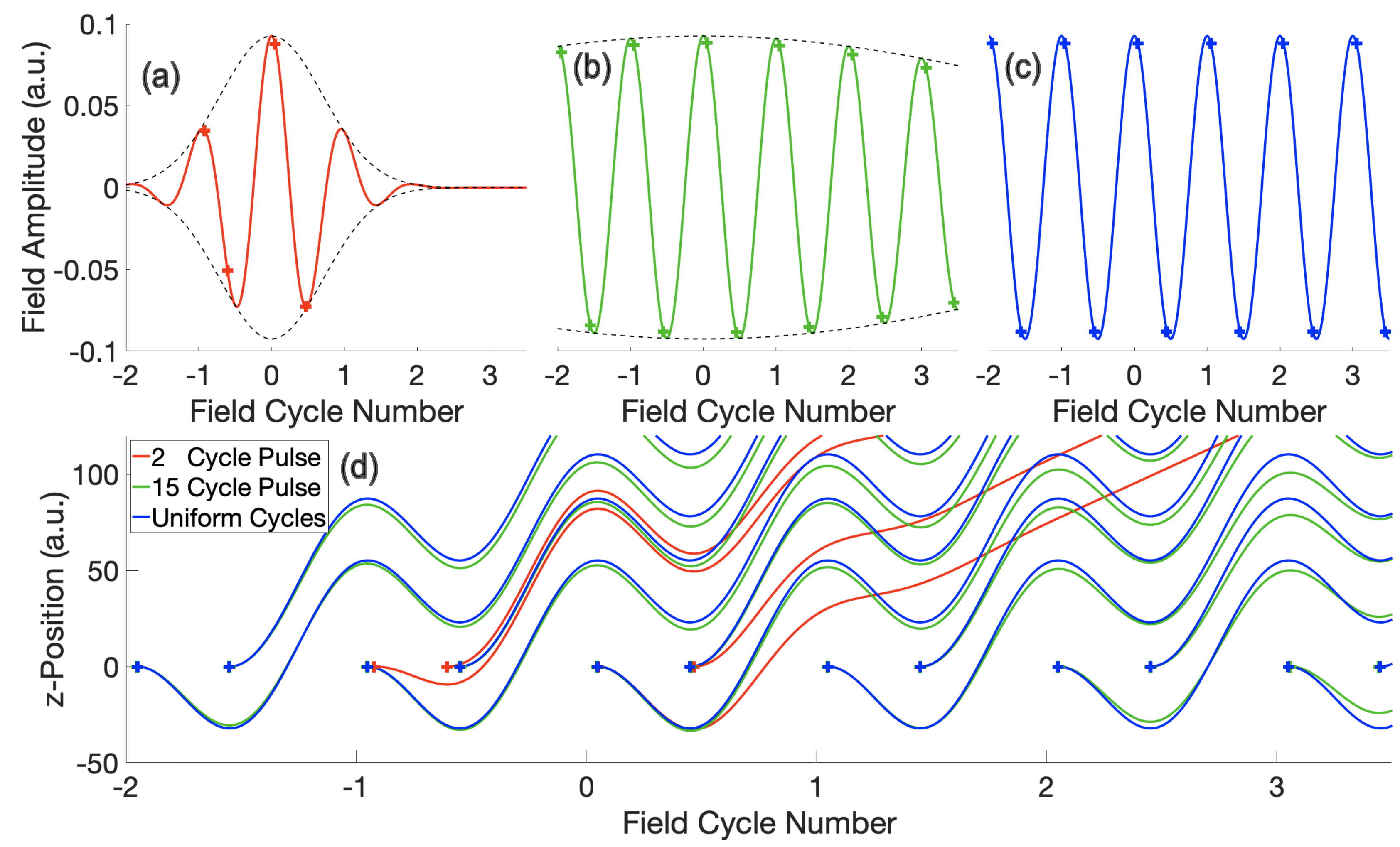}
	\caption{Comparison of the photoelectron trajectories produced by three distinct 800~nm laser pulses. (a) and (b) show the electric field of a 2-cycle ($\sim$ 5.3~fs) pulse and a 15-cycle ($\sim$ 40~fs) pulse respectively. The carrier envelope is indicated by the dashed black line for each. (c) shows a uniform cycle laser field pulse. In (a), (b), and (c) the ionization times are plotted as crosses for an electron which ends up at ($p_\parallel$, $p_\perp$) = (0.5, 0) on the detector (see main text). (d) displays the electron trajectories propagating in each of the laser fields above. Each trajectory begins at the time of ionization indicated in plots (a), (b), or (c), and arrives at the same final momentum (0.5, 0). 
	The 15-cycle pulse much better reproduces the trajectories generated by the uniform-cycle pulse, as the carrier envelope parameter has less of an impact.
	For simplicity, no rescattering trajectories are shown.}
	\label{fig:Trajectories}
\end{figure*}

A primary way to isolate individual features is to compare the measured spectrum to quantum SFI calculations. In the strong-field regime, electron spectra are largely determined by the phase of the laser field at the moment of photoionization, and the subsequent evolution in the field. This is known as the strong-field approximation (SFA) \cite{faisal_multiple_1973,reiss_semiclassical_1970}.
Of particular interest for the characterization of the atomic or molecular target are interference patterns produced by distinct electron trajectory pairs ending up at the same final momentum whose ionization times are within a single laser field cycle of each other. These features are labeled as sub-cycle, whereas features produced by electron trajectory pairs ionized at least one full field cycle apart we label as inter-cycle. 
Referring to Fig. \ref{fig:Trajectories} (d), electron trajectories resulting from different field cycles are roughly periodic, and change very little between field cycles. 
Because interferences due to inter-cycle trajectory pairs are dependent primarily on the shape of the laser pulse rather than any property of the target, they do not encode any new information about the target, but still produce additional undesired features in the momentum distribution \cite{faria_it_2020}.
Importantly, angle-resolved SFI spectra produced by a multi-cycle laser pulse contain structures produced from both sub-cycle and inter-cycle interfering trajectory pairs.
To mitigate this, calculations of sub-cycle features often assume strong-field conditions that are practically unattainable, such as a few uniform strong-field cycles that turn on and off instantly \cite{faria_it_2020, maxwell_coulomb-free_2018}. Experiments cannot mimic this.
Instead, realizable ultrashort 1- or 2-cycle laser pulses include a time-varying field envelope, which gives rise to a carrier envelope phase (CEP) parameter that governs the electron dynamics but is absent from calculations that assume uniform cycles (See Fig. \ref{fig:Trajectories}) \cite{wittmann_single-shot_2009,kling_imaging_2008}. These pulses generate spectra that do not match sub-cycle calculations, despite having eliminated inter-cycle features.
Thus to complement the ultrashort approach and more closely produce experimental spectra comparable to these few-cycle calculations, we require a different approach.

\begin{figure*}
	\centering
	\includegraphics[width=1\linewidth]{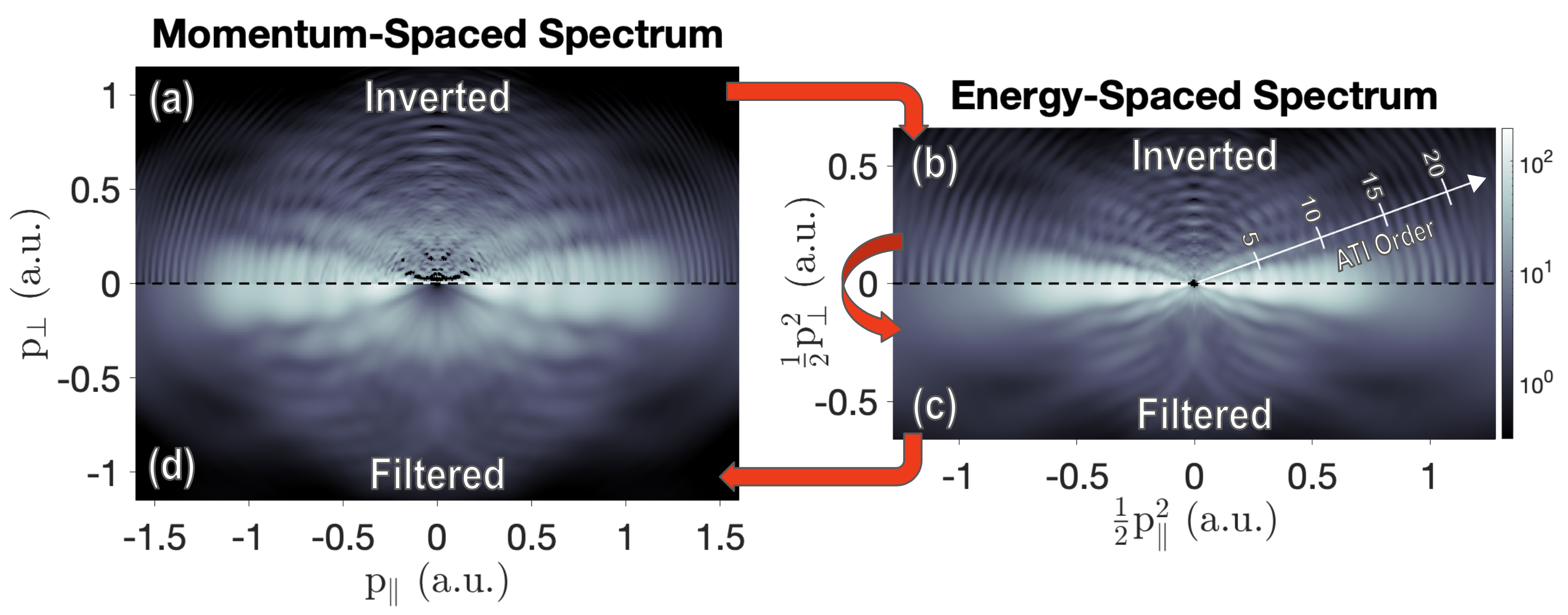}
	\caption{The application of the time-filtering technique to our argon spectrum. Here, we have zoomed into the spectrum to emphasize the region within 2U$_{\mathrm{p}}$ where the filtering has the greatest effect. The red arrows outline the order of the data processing, and the colorscale indicates the normalized photoelectron yield. (a) The top half of the image shows the reconstructed $p_\parallel-p_\perp$ momentum slice from polar onion-peeling \cite{roberts_toward_2009}. (b) Transforming the onion-peeled spectrum to a polar energy plot yields the top half of the image. In this representation the ATI rings are equally spaced, the radial axis counts the ATI order, and the x- and y-axes label the projection of the squared components of momenta as indicated. (c) The bottom half shows the filtered energy-spaced spectrum as detailed in the text on the same scale. (d) The bottom half shows the filtered momentum slice on the same scale as the original spectrum. Importantly for (c) and (d) we note the time-filtering has only removed the ATI rings, and all other visible structures remain unaffected.}
	\label{fig:VMI_Processing}
\end{figure*}

We are interested in sub-cycle interfering electron trajectory pairs in SFI generated by a standard 30-50~fs multi-cycle laser pulse from a commercial Ti:Sapphire laser system. This requires identifying and distinguishing the key features of inter-cycle dynamics and sub-cycle dynamics present in multi-cycle spectra. In the tunneling ionization regime, the primary inter-cycle features appear as a comb of ATI peaks, which are equally spaced by the energy of the driving laser photons \cite{freeman_above-threshold_1987,marchenko_wavelength_2010}. Under ordinary conditions, this comb dominates the momentum spectrum, and frequently obscures patterns generated by sub-cycle dynamics. This is most noticeable in the direct ionization regime of the spectrum below 2U$_{\mathrm{p}}$, where U$_{\mathrm{p}}$ is the ponderomotive energy of a free electron in the laser field \cite{bucksbaum_role_1987}. 
Here, holographic features resulting from the interference of electron trajectories after ionization are prevalent \cite{faria_it_2020, huismans_time-resolved_2011}. These sub-cycle features would be far more visible if the ATI contribution to the spectrum could be filtered out.

In this paper, we present a time-filtering technique that effectively eliminates interference patterns from trajectory pairs ionized at least one field cycle apart, which removes the energy-periodic background ATI comb in multi-cycle photoelectron momentum spectra. This suppresses the inter-cycle contributions to the spectrum, thus leaving only the sub-cycle dynamics. Importantly, since we use a multi-cycle laser pulse, the resultant spectrum closely resembles that due to a single cycle of a steady-state laser field (See Fig. \ref{fig:Trajectories} (b-d)), which models common SFI calculations more closely \cite{maxwell_coulomb-free_2018, faria_it_2020, smirnova_strong-field_2009, smirnova_high_2009, yuan_exploring_2021, pisanty_slalom_2016}. 
Furthermore, the time-filtering technique is broadly applicable to angle-resolved SFI electron spectra and is independent of experimental parameters, making it a simple tool for extracting sub-cycle structures from multi-cycle data. To illustrate the time-filtering, we demonstrate it on a high-fidelity VMI spectrum of argon gas photoionized with an approximately 40~fs, 800~nm commercial Ti:sapphire laser at a peak intensity of 200~TW/cm$^2$. This time-filtering method reveals new features of sub-cycle ionization that should be of interest in the study of holography in SFI. All quantities are in atomic units, unless otherwise stated.

The raw VMI detector image requires several layers of processing to reach the effective single-cycle spectrum. The steps are outlined below and in Fig.~\ref{fig:VMI_Processing}. The two-dimensional VMI raw detector image is first calibrated to measure the transverse momentum of the electrons striking it \cite{dahl_simion_1995}. The resulting image is a projection on ($p_\parallel$, $p_\perp$) of a three-dimensional Newton sphere of electron momenta $(p_\parallel$, $p_\perp$, $\phi)$ \cite{eppink_velocity_1997}. Here $p_\parallel$ is the momentum magnitude parallel to the polarization axis of the ionizing field, $p_{\perp}$ is the momentum magnitude transverse to polarization axis, and $\phi$ is the azimuthal angle about the polarization axis. Since photoionization with linearly polarized light is cylindrically symmetric about the polarization ($\parallel$) axis, reconstructing just the $p_\parallel-p_\perp$ slice of the Newton Sphere from its full projection is sufficient to display the full three-dimensional information \cite{reid_photoelectron_2003}. 

Reconstructing this momentum slice begins by symmetrizing the VMI data. Each raw spectrum is centered and rotated, and four-quadrant symmetry is imposed by averaging each pixel in the raw image with its counterparts in the other three quadrants. 

\begin{figure}
	\centering
	\includegraphics[width=1\linewidth]{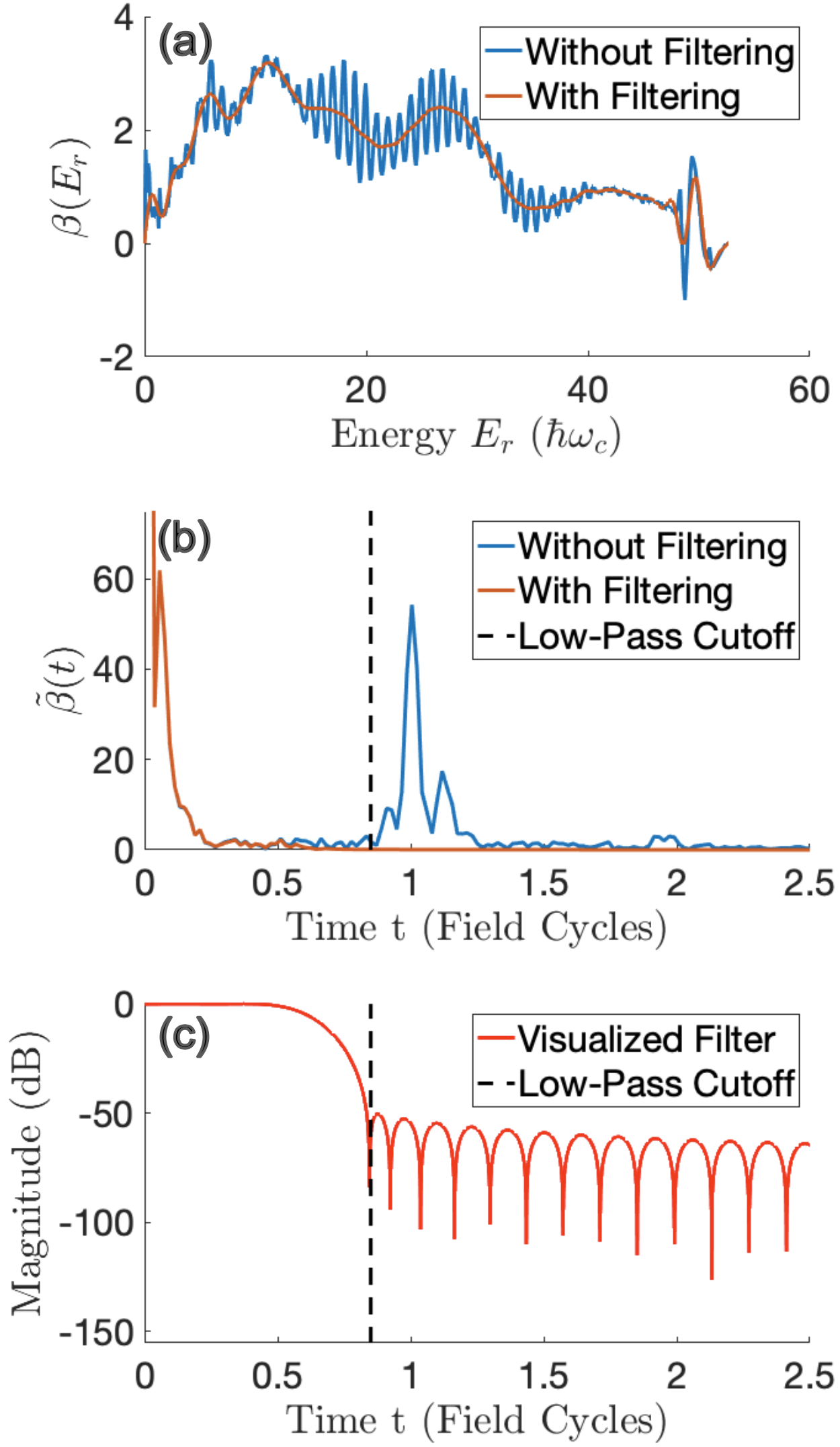}
	\caption{(a) The $\beta_6(E_r)$ parameter of the spectrum in Fig.\ref{fig:VMI_Processing} presented in units of the energy of a single photon at the central laser frequency, $\hbar \omega_c$. ($\beta_6(E_r)$ is depicted as an example of the filtering method.  The other orders are analyzed in an identical way.) Shown are the same parameter before (blue) and after (orange) time-filtering. Prior to filtering we see the evenly spaced ATI comb quite prominently. After we apply the time filter we see that the filtering only removes the ATI comb, and does not disrupt the underlying shape of the parameter. (b) The Fourier transform of $\beta_6(E_r)$. The large spike at $t=1$ field cycle is due to the ATI comb. 
	A low-pass filter with a cut-off (black dashed line) just below this peak removes only the ATI comb. Below the cut-off, the filtered and unfiltered $\tilde{\beta}_6(t)$ match very closely. This is achieved by designing a filter that is maximally flat in the pass band. (c) This shows the magnitude response of the low-pass filter applied to each of the $\beta_n(E_r)$ parameter. Note the smoothness of the magnitude response in the pass band below the cut-off which helps prevent nonphysical artifacts from appearing in the data as a result of the filtering.}
	\label{fig:Beta_Parameters}
\end{figure}

The projection is inverted through the procedure of polar onion-peeling \cite{roberts_toward_2009}. Beginning from the outermost radius of the raw data, we construct the Legendre decomposition least squares fit to the photoelectron angular distribution (PAD) at that radius using the Legendre basis polynomials of even orders $P_n(\cos(\theta))$. 

\begin{equation}
    I(\theta, p_r) = C_0(p_r)\sum_{n_{\mathrm{even}}} \beta_n(p_r)P_n(\cos(\theta))
    \label{eqn:L_Decomposition}
\end{equation}

Here, $\theta$ is the polar angle about the center of the 2-dimensional slice and $I(\theta, p_r)$ is the PAD fit at the momentum radius $p_r$. The anisotropy parameters $\beta_n(p_r)$ are defined as $\beta_n(p_r) = \frac{C_n(p_r)}{C_0(p_r)}$ where $C_n(p_r)$ is the nth Legendre coefficient for the decomposition at $p_r$. In our decomposition, the maximum Legendre order is set to 42, well beyond the required angular resolution for holographic predictions in SFI \cite{faria_it_2020}. We then convolve this fit about the polarization axis and project the resulting spherical shell back to the image plane. This generates the projection contribution from that spherical shell of the Newton sphere. Subtracting this contribution from the spectrum ``peels'' this shell of the Newton sphere off the spectrum. Repeating this procedure for successively smaller radii yields the fully inverted spectrum.

This process generates the momentum slice $p_\parallel-p_\perp$ as a superposition of Legendre polynomials whose anisotropy parameters, $\beta_n(p_r)$, are one-dimensional functions of $p_r$. All of the information in the two-dimensional momentum distribution is now encoded in a one-dimensional $\beta_n(p_r)$ parameter for each included order $n$ \cite{reid_photoelectron_2003}. 


At this point in the analysis the sub-cycle features are obscured by highly prominent ring structures of inter-cycle ATI interference \cite{freeman_above-threshold_1987}. The ATI appears as a comb of peaks in each order of $\beta_n(p_r)$ spaced by the photon energy of the ionizing laser, so we transform the spectral grid from momentum to energy by directly resampling $\beta_n(p_r) \rightarrow \beta_n(E_r)$. The result is shown in panel (b) of Fig.~\ref{fig:VMI_Processing}. Since in this spacing the peaks are periodic, we can take the Fourier transforms of each $\beta_n(E_r)$ (Fig.~\ref{fig:Beta_Parameters} (b)) which contain a sharp peak due to the ATI. The reciprocal space of energy is time, and in this case the times are the time differences between two interfering electron trajectories, as highlighted in Fig. \ref{fig:Trajectories}. We see a large spike at a time difference of one field cycle, which is consistent with the ATI forming due to the interference of similar trajectories from different laser cycles. 

\begin{figure}
	\centering
	\includegraphics[width=1\linewidth]{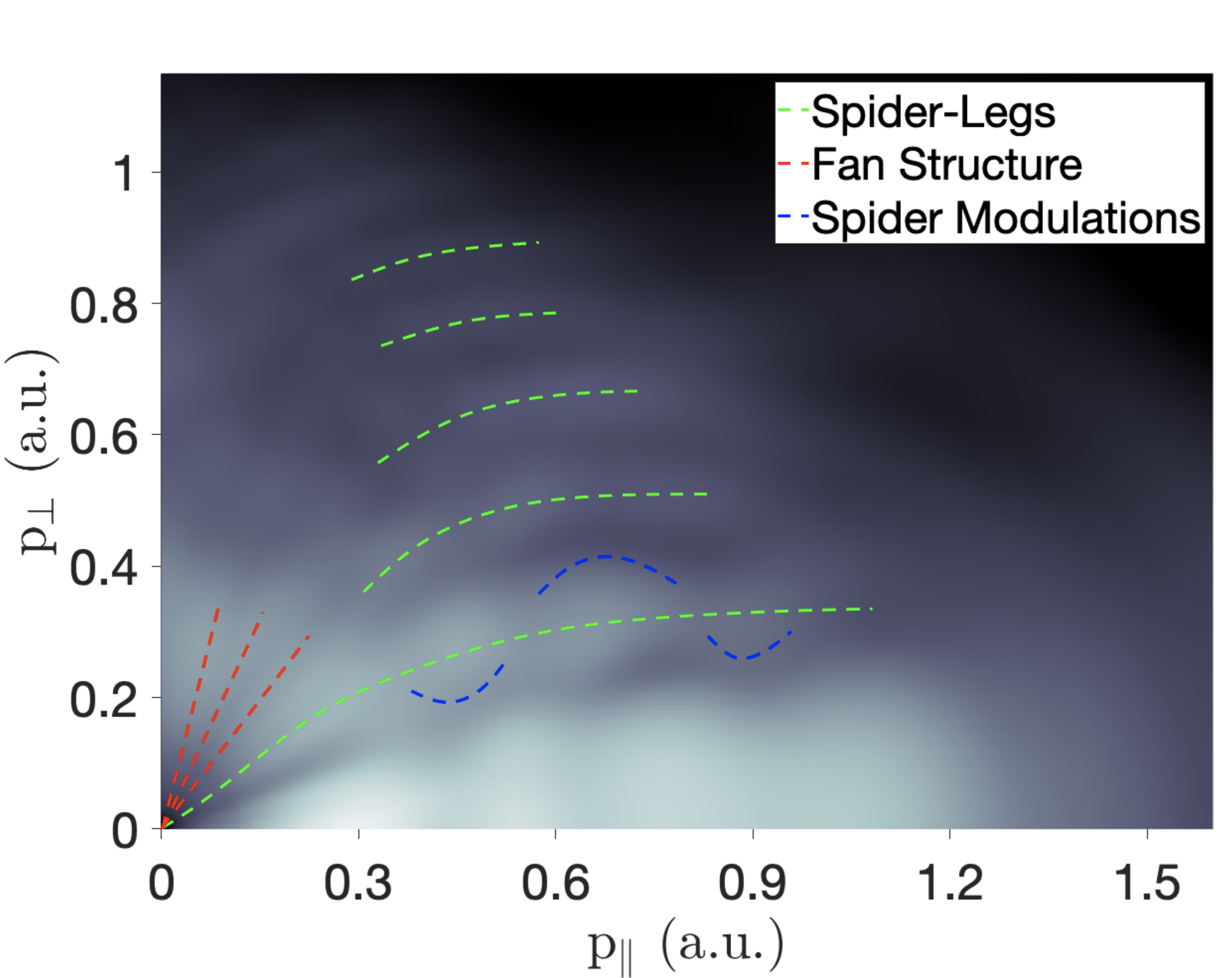}
	\caption{Magnified image of a single quadrant of the filtered spectrum shown in Fig.~\ref{fig:VMI_Processing} (d) on the same logarithmic color scale. The spider-leg structure, fan structure, and previously unexplored wiggling modulations along the spider-leg structures are highlighted by dashed lines as shown. Periodic minima along the polarization axis can also be seen. These unexplored modulations are normally obscured by the ATI rings.}
	\label{fig:Filtered_Spectrum}
\end{figure}

In this way we identify that sub-cycle interference patterns are encoded in $\tilde{\beta}_n(t)$ as all structure below the time difference of one field cycle. This suggests that in order to suppress the effect of inter-cycle interference features such as ATI rings we can Fourier filter $\beta_n(E_r)$. We construct a zero-phase, finite impulse response low-pass filter, with a cut-off just below one field cycle to suppress the ATI peak and further inter-cycle interference features, without affecting any sub-cycle structures in the spectrum. A Kaiser window filter is selected to make the response of the signal maximally flat in the pass band (See Fig.~\ref{fig:Beta_Parameters} (c)). The result of this filtering can be seen in Fig. \ref{fig:VMI_Processing} (c). Transforming back to momentum spacing returns the original momentum grid, and the image can be reproduced according to Eq.~\ref{eqn:L_Decomposition}. The final filtered spectrum does not contain the ATI interference pattern, but does preserve the features that are caused by sub-cycle interference in field ionization and rescattering, as shown in Fig.~\ref{fig:Filtered_Spectrum}.

Fig.~\ref{fig:Filtered_Spectrum} resolves previously unexplored interference features in the holographic regime of the spectrum that are normally obscured by the ATI rings. Most strikingly, along the spider-leg structures, which have been studied extensively in earlier literature \cite{huismans_time-resolved_2011, hickstein_direct_2012, gong_pathway-resolved_2016, moller_off-axis_2014, li_fine_2014, korneev_interference_2012}, we identify new subtle wiggling modulations that have not been examined previously. This is most clearly visible in the first order spider leg in the range $0.4<p_\parallel<1$ and $0.15<p_\perp<0.4$, though the same modulations can be seen with increasing difficulty at higher order legs. Additionally, we can resolve periodic minima along the polarization axis and each spider leg. Clear observations of these holographic features can help validate strong-field calculations and lead to a more complete understanding of the strong-field driven dynamics that generate them.

Although we have only demonstrated the time-filtering procedure using a single VMI spectrum, the idea behind it is quite general and broadly applicable to momentum-resolved electron spectra collected in other ways. The key takeaway is that inter-cycle interference patterns can be isolated and extracted through the Fourier transform of the yield of electrons versus the electron energy, which can be determined from the momentum spectrum. This Fourier transform describes the time difference between the ionization times of interfering electron trajectories, which should hold true for any method of momentum-resolved electron collection. Here we outlined polar onion-peeling as our inversion method, which uses the Legendre basis to construct the PAD fits before inverting to produce the $p_\parallel-p_\perp$ momentum slice. This too can be generalized. Polar onion-peeling is advantageous for the time-filtering as it produces one-dimensional, polar anisotropy parameters which can each be filtered individually to preserve the angular resolution of the spectrum. However, this can be achieved for any angularly-resolved spectrum by constructing PAD fits after any desired inversion method, and then filtering the produced coefficients as outlined previously in the text.


In conclusion, we have presented a procedure to filter out the ATI comb and other inter-cycle features in a strong-field ionization spectrum, to more clearly reveal the underlying sub-cycle features of the strong-field ionization process. This time-filtering is not simply an aesthetic change. The filtered momentum spectrum shown in Fig.~\ref{fig:Filtered_Spectrum} contains holographic features from sub-cycle trajectory interferences that are normally obscured by the ATI rings \cite{faria_it_2020}. These sub-cycle interferences revealed by the filtering merit significant further study and should help to validate holographic models in strong-field ionization.

\begin{acknowledgments}

We would like to thank J. Cryan for his invaluable help in the laboratory in getting the experimental apparatus up and running. We are also grateful to A. Maxwell and C. Faria for their very illuminating discussions on the nature of sub-cycle trajectory interferences in strong field ionization and how they display in photoelectron momentum distributions.

This work is supported by the U.S. Department of Energy, Office of Science, Basic Energy Sciences (BES), Chemical Sciences, Geosciences, and Biosciences Division, AMOS Program.

\end{acknowledgments}



\bibliography{main}

\end{document}